\documentclass[draftcls, onecolumn, 12pt]{IEEEtran}
\usepackage{amsmath,graphicx}
\usepackage{amssymb}
\usepackage{subfigure}
\usepackage{footmisc}
\usepackage{lipsum}
\usepackage{mathtools}
\usepackage{cuted}
\usepackage{esint}
\usepackage{multicol}
\usepackage{cite}
\usepackage{xcolor}
\makeatletter
\@for\next:={int,iint,iiint,iiiint,dotsint,oint,oiint,sqint,sqiint,
  ointctrclockwise,ointclockwise,varointclockwise,varointctrclockwise,
  fint,varoiint,landupint,landdownint}\do{%
    \expandafter\edef\csname\next\endcsname{
      \noexpand\DOTSI
      \expandafter\noexpand\csname\next op\endcsname
      \noexpand\ilimits@
    }%
  }
\makeatother
\begin{document}

\title{Reliability Analysis of Large Intelligent Surfaces (LISs): Rate Distribution and Outage Probability}

\author{Minchae~Jung,~\IEEEmembership{Member,~IEEE}, Walid~Saad,~\IEEEmembership{Fellow,~IEEE}, \\
Youngrok~Jang, Gyuyeol~Kong,
and Sooyong~Choi,~\IEEEmembership{Member,~IEEE} 
\thanks{
{\hspace{-9pt}M. Jung, Y. Jang, G. Kong, and S. Choi (corresponding author) are with School of Electrical Electronic Engineering, Yonsei University, Seoul 03722, Korea (e-mail: csyong@yonsei.ac.kr).}

{W. Saad is with Wireless@VT, Department of Electrical and Computer Engineering, Virginia Tech, Blacksburg, VA 24061 USA (e-mail: walids@vt.edu).}

This research was supported by Basic Science Research Program through the National Research Foundation of Korea (NRF) funded by the Ministry of Education (NRF-2016R1A6A3A11936259)
and by the U.S. National Science Foundation under Grants CNS-1836802 and OAC-1638283.
}
}

\markboth{IEEE WIRELESS COMMUNICATIONS LETTERS,~Vol.~0, No.~0, XX~201X}%
{Shell \MakeLowercase{\textit{et al.}}: Bare Demo of IEEEtran.cls for Journals}

\maketitle
\begin{abstract}
Large intelligent surfaces (LISs) have been recently proposed as an effective wireless communication solution that can leverage antenna arrays deployed on the entirety of man-made structures such as walls.
An LIS can provide space-intensive and reliable communication, enabling the desired wireless channel to exhibit a perfect line-of-sight.
However, the outage probability of LIS, which is an important performance metric to evaluate the system reliability, remains uncharacterized.
In this paper, the distribution of uplink sum-rate is asymptotically analyzed for an LIS system.
Given the derived asymptotic distribution, the outage probability is derived for the considered LIS system.
Simulation results show that the results of the proposed asymptotic analyses are in close agreement to the exact mutual information in the presence of a large number of antennas and devices.
\end{abstract}



\IEEEpeerreviewmaketitle
\section{Introduction}
\IEEEPARstart{T}{he} demand for wireless connectivity has been growing exponentially in recent years, mainly driven by the variety of upcoming Internet of Things (IoT) applications, such as sensors, vehicles, and drones \cite{ref.Saad2019vision,ref.Mozaffari2019beyond}.
To support this demand for wireless connectivity for IoT services without additional radio resources, the concept of large intelligent surfaces (LISs) has been newly proposed to exploit the fact that 
man-made structures, such as buildings and walls, can be made electromagnetically active and used for wireless transmission \cite{ref.Hu2018data,ref.Hu2018HWI,ref.Jung2018lisul,ref.Huang2018ee_res,ref.Huang2019indoor}.
An LIS system enables the desired channels to become line-of-sight (LOS) channels,
resulting in more reliable and space-intensive communications compared to conventional massive multiple-input multiple-output (MIMO) systems.\footnote{For an overview on how LIS differs from massive MIMO, see \cite{ref.Hu2018data,ref.Hu2018HWI,ref.Jung2018lisul,ref.Huang2018ee_res,ref.Huang2019indoor}.}
Effectively using LISs requires addressing many challenges such as obtaining their achievable rate, evaluating the system reliability, and analyzing the channel hardening effect.

Prior art \cite{ref.Hu2018data,ref.Hu2018HWI,ref.Jung2018lisul,ref.Huang2018ee_res,ref.Huang2019indoor} has studied some of these challenges.
In particular, the work in \cite{ref.Hu2018data} studied uplink data rate of the matched filter (MF) and derived the performance of the optimal receiver in an LIS system.
In \cite{ref.Hu2018HWI} and \cite{ref.Jung2018lisul}, the authors analyzed the uplink data rates to evaluate LIS performance while considering, respectively, hardware impairments and channel estimation errors.
Moreover, the work in \cite{ref.Jung2018lisul} verified the occurrence of the channel hardening effect theoretically.
{
Meanwhile, the authors in \cite{ref.Huang2018ee_res} and \cite{ref.Huang2019indoor} proposed the use of LIS as a reconfigurable scatterer
that reflects transmitted signals and developed LIS phase shifter that maximize energy efficiency and received signal strength, respectively.
However, these recent works \cite{ref.Hu2018data,ref.Hu2018HWI,ref.Jung2018lisul,ref.Huang2018ee_res,ref.Huang2019indoor} have not investigated the distribution of the actual data rate which is necessary to analyze the outage probability of an LIS.}
Since an accurate estimation of the outage probability enables us to operate an LIS network reliably,
it is necessary to analyze this outage probability in an LIS system serving a large number of devices.

{The main contribution of this paper is to derive an exact closed-form expression for the outage probability 
and verify the reliability of an LIS system in terms of the outage probability.\footnote{
{Note that those results cannot be obtained directly from \cite{ref.Jung2018lisul}.}}}
Given an LIS area serving multiple devices, we analyze the asymptotic distribution of the sum-rate, relying on the Lyapunov central limit theorem (CLT) in the presence of a large number of antennas and devices.
This approximation allows for accurate estimations of the outage probability without the need for a large number of simulations, and it enables us to verify the reliability of an LIS system.
Simulation results show that an LIS can provide reliable communication regardless of signal-to-noise-ratio (SNR) given that
the outage probability keeps constant for varying SNR when the LIS is equipped
with a large number of antennas and devices.




\section{System Model}
Consider an uplink LIS system in which a large number of IoT devices are connected to the LIS's large array and a two-dimensional LIS is deployed on the horizontal plane.
The LIS consists of $K$ LIS units, each of which occupies a subarea of the entire LIS
and has a square shape with limited area of $2L \times 2L$ serving a single-antenna device. 
{
We assume that the LIS has its own signal processing unit for estimating channel and detecting any data signal, as in \cite{ref.Hu2018data,ref.Hu2018HWI,ref.Jung2018lisul}.
}
A large number of antennas, $M$, are deployed on the surface of the LIS unit in rectangular lattice form with $\Delta L$ spacing, centered on the $(x, y)$ coordinates of the corresponding device.
Given that the location of device $k$ is $({x_k},{y_k},{z_k})$, antenna $m$ of LIS unit $k$ is placed at $({x^{\rm{LIS}}_{km}},{y^{\rm{LIS}}_{km}},0)$ where $x_{km}^{{\rm{LIS}}} \in [ {{x_k} - L,{x_k} + L} ]$ and $y_{km}^{{\rm{LIS}}} \in [ {{y_k} - L,{y_k} + L} ]$.
The LIS units may overlap depending on the location of their corresponding devices and this results in severe performance degradation. 
To prevent this problem, 
we assume that the devices with partially overlapping LIS units are allocated on orthogonal resources, resulting in an LIS composed of $K$ non-overlapping LIS units.
Moreover, we assume that each device controls its uplink transmit power toward the center of its LIS unit according to a target SNR. 

\subsection{Wireless Channel Model}
{Note that the entire LIS environment is active during wireless communication and the signal from the NLOS path can be negligible compared to the LOS signal, as proved in \cite{ref.Jung2018lisul}.
Hence, we consider the LIS channel ${{\boldsymbol{h}}_{kk}} \in \mathbb{C}{^M}$ between device $k$ and LIS unit $k$ as a LOS path defined as}
${{\boldsymbol{h}}_{kk}}{=}{[ {\beta _{k1}^{\rm{L}}{h_{kk1}}, \cdots ,\beta _{kM}^{\rm{L}}{h_{kkM}}} ]^{\rm{T}}}$,
where $\beta _{km}^{\rm{L}} {=} \alpha _{kkm}^{\rm{L}}l_{kkm}^{\rm{L}}$ and ${h_{kkm}} {=} \exp \left( { - j2\pi {d_{kkm}}/\lambda } \right)$ denote a LOS channel gain and state, respectively, between device $k$ and antenna $m$ of LIS unit $k$. 
${\alpha_{kkm}^{\rm{L}}} {=} \sqrt { z_k/d_{kkm}} $ and $l_{kkm}^{\rm{L}} {=} 1/\sqrt {4\pi d_{kkm}^2}$ represent, respectively, the antenna gain and path loss attenuation,
where $d_{kkm}$ and $\lambda$ denote the distance from device $k$ to antenna $m$ of LIS unit $k$
and the wavelength of a signal, respectively.

Given a Rician factor $\kappa _{jk}$, we consider the interference channel ${{\boldsymbol{h}}_{jk}} \in \mathbb{C}{^M}$ between device $j$ and LIS unit $k$ as a Rician fading channel, given by
${{\boldsymbol{h}}_{jk}} = \sqrt {\frac{{{\kappa _{jk}}}}{{{\kappa _{jk}} + 1}}} {\boldsymbol{h}}_{jk}^{\rm{L}} + \sqrt {\frac{1}{{{\kappa _{jk}} + 1}}} {\boldsymbol{h}}_{jk}^{{\rm{NL}}}$, 
where ${\boldsymbol{h}}_{jk}^{\rm{L}} ={[ {\beta _{j1}^{\rm{L}}{h_{jk1}}, \cdots ,\beta _{jM}^{\rm{L}}{h_{jkM}}} ]^{\rm{T}}}$ and ${\boldsymbol{h}}_{jk}^{\rm{NL}} =  {\boldsymbol{R}}_{jk}^{1/2}{{\boldsymbol{g}}_{jk}}$ denote the deterministic LOS and the correlated NLOS component, respectively.
Given $P$ dominant paths among all NLOS paths, we define ${{\boldsymbol{R}}_{jk}} \in \mathbb{C}{^{M \times P}}$ and ${{\boldsymbol{g}}_{jk}} = {[ {{g_{jk1}}, \cdots ,{g_{jkP}}} ]^{\rm{T}}} \sim \mathcal{CN}\left( {{\boldsymbol{0}},{{\boldsymbol{I}}_P}} \right)$ as the deterministic correlation matrix and an independent fading channel between device $j$ and LIS unit $k$, respectively.
Since a two-dimensional LIS is deployed on the $xy$-plane, 
we can model it as a uniform planar array \cite{ref.Song2017common}.
Then, the correlation matrix can be defined as
${\boldsymbol{R}}_{jk}^{1/2} = {\boldsymbol{l}}_{jk}^{{\rm{NL}}}{{\boldsymbol{D}}_{jk}}$, where 
${\boldsymbol{l}}_{jk}^{{\rm{NL}}} = {\rm{diag}}( {l_{jk1}^{{\rm{NL}}}  \cdots ,l_{jkM}^{{\rm{NL}}}} )$ is a diagonal matrix that includes the path loss attenuation $l_{jkm}^{{\rm{NL}}} = d_{jkm}^{ - {\beta _{{\rm{PL}}}}/2}$ with a path loss exponent ${\beta _{{\rm{PL}}}}$ and
${{\boldsymbol{D}}_{jk}} = [ {\alpha _{jk1}^{{\rm{NL}}}{\boldsymbol{d}}( {\phi _{jk1}^{\rm{v}},\phi _{jk1}^{\rm{h}}} ), \cdots ,\alpha _{jkP}^{{\rm{NL}}}{\boldsymbol{d}}( {\phi _{jkP}^{\rm{v}},\phi _{jkP}^{\rm{h}}} )} ]$.
The term ${\boldsymbol{d}}( {\phi _{jkp}^{\rm{v}},\phi _{jkp}^{\rm{h}}} )$ represents the NLOS path $p$ at given angles $( {\phi _{jkp}^{\rm{v}},\phi _{jkp}^{\rm{h}}} )$, defined as
${\boldsymbol{d}}( {\phi _{jkp}^{\rm{v}},\phi _{jkp}^{\rm{h}}} ) = \frac{1}{{\sqrt M }}{{\boldsymbol{d}}_{\rm{v}}}( {\phi _{jkp}^{\rm{v}}} ) \otimes {{\boldsymbol{d}}_{\rm{h}}}( {\phi _{jkp}^{\rm{h}}} ),$
where
$\otimes$ is the Kronecker product and
\begin{align}
{{\boldsymbol{d}}_{\rm{v}}}( {\phi _{jkp}^{\rm{v}}} ) &={\left[ {1,{\rm{ }}{e^{j\frac{{2\pi \Delta L}}{\lambda }\phi _{jkp}^{\rm{v}}}}, \cdots ,{\rm{ }}{e^{j\frac{{2\pi \Delta L}}{\lambda }\left( {\sqrt M  - 1} \right)\phi _{jkp}^{\rm{v}}}}} \right]^{\rm{T}}},\nonumber\\
{{\boldsymbol{d}}_{\rm{h}}}( {\phi _{jkp}^{\rm{h}}} ) &={\left[ {1,{\rm{ }}{e^{j\frac{{2\pi \Delta L}}{\lambda }\phi _{jkp}^{\rm{h}}}}, \cdots ,{\rm{ }}{e^{j\frac{{2\pi \Delta L}}{\lambda }\left( {\sqrt M  - 1} \right)\phi _{jkp}^{\rm{h}}}}} \right]^{\rm{T}}}.\nonumber
\end{align}

\noindent Here, $\phi _{jkp}^{\rm{v}} = \sin \theta _{jkp}^{\rm{v}}$ and $\phi _{jkp}^{\rm{h}} = \sin \theta _{jkp}^{\rm{h}}\cos \theta _{jkp}^{\rm{h}}$ 
when the elevation and azimuth angles between device $j$ and LIS unit $k$ at path $p$ are $\theta _{jkp}^{\rm{v}}$ and $\theta _{jkp}^{\rm{h}}$, respectively. 
Further, $\alpha _{jkp}^{{\rm{NL}}} = \sqrt {\cos \theta _{jkp}^{\rm{v}}\cos \theta _{jkp}^{\rm{h}}}$ denotes the antenna gain at path $p$ with ${\theta _{jkp}} \in [ { - \frac{\pi }{2},\frac{\pi }{2}} ]$ and ${\theta _{jkp}} \in \{ {\theta _{jkp}^{\rm{v}},\theta _{jkp}^{\rm{h}}} \}$.

\subsection{Uplink Data Rate}
The received signal from all devices at LIS unit $k$ is obtained as
\begin{equation}
{{\boldsymbol{y}}_k} = \sqrt {{\rho_k}} {{\boldsymbol{h}}_{kk}}{x_k} + \sum\nolimits_{j \ne k}^K {\sqrt {{\rho_j}} {{\boldsymbol{h}}_{jk}}{x_j}}  + {{\boldsymbol{n}}_k}, \nonumber
\end{equation}
where ${x_k}$ and ${x_j}$ are uplink signals of device $k$ and $j$, respectively, ${\rho_k}$ and ${\rho_j}$ are their transmit SNRs, and ${{\boldsymbol{n}}_k} \in\mathbb{C} {^M} \sim \mathcal{CN}\left( {{\boldsymbol{0}},{{\boldsymbol{I}}_M}} \right)$ is noise vector.
Given a linear receiver ${\boldsymbol{f}}_k^{\rm{H}}$ at LIS unit $k$, we have
\begin{equation}
{\boldsymbol{f}}_k^{\rm{H}}{{\boldsymbol{y}}_k} = \sqrt {{\rho_k}} {\boldsymbol{f}}_k^{\rm{H}}{{\boldsymbol{h}}_{kk}}{x_k} + \sum\nolimits_{j \ne k}^K {\sqrt {{\rho_j}} {\boldsymbol{f}}_k^{\rm{H}}{{\boldsymbol{h}}_{jk}}{x_j}}  + {\boldsymbol{f}}_k^{\rm{H}}{{\boldsymbol{n}}_k}. \label{eq.1}
\end{equation}
Given an MF receiver with imperfect channel estimation results from a least square estimator, we have ${{\boldsymbol{f}}_k} = {{\boldsymbol{h}}_{kk}} + \sqrt{{\tau_k^2}/\left({1-\tau_k^2}\right)}{{\boldsymbol{e}}_k}$ where ${{\boldsymbol{e}}_k}$ is the error vector uncorrelated with ${{\boldsymbol{n}}}_{k}$ \cite{ref.Jung2018lisul}.
From (\ref{eq.1}), the received signal-to-interference-plus-noise ratio (SINR) at LIS unit $k$ will be
${\gamma _k} = {{{\rho _k}{S_k}\left( {1 - \tau_k^2} \right)}}/{{{I_k}}}$,
where 
${S_k} = {{\left| {{{\boldsymbol{h}}_{kk}}} \right|}^4}$ and
${I_k} = {{{\rho _k}\tau_k^2{X_k} + \sum\nolimits_{j \ne k}^K {{\rho _j}{Y_{jk}}}  + {Z_k}}}$.
Here,
${X_k}$, ${Y_{jk}}$, and ${Z_k}$ denote ${{| {{\boldsymbol{e}}_k^{\rm{H}}{{\boldsymbol{h}}_{kk}}} |}^2}$,
 ${{| {\sqrt {1 - \tau_k^2} {\boldsymbol{h}}_{kk}^{\rm{H}}{{\boldsymbol{h}}_{jk}} + {\tau_k}{\boldsymbol{e}}_k^{\rm{H}}{{\boldsymbol{h}}_{jk}}} |}^2}$, and
${{| {\sqrt {1 - \tau_k^2} {\boldsymbol{h}}_{kk}^{\rm{H}} + {\tau_k}{\boldsymbol{e}}_k^{\rm{H}}} |}^2}$, respectively.
From the received SINR at LIS unit $k$, the uplink sum-rate is obtained by 
${R} = \sum\nolimits_{k = 1}^K {R_k}$, where ${R_k} = \log \left( {1 + {\gamma _k}} \right)$ is instantaneous rate for device $k$.
Given this sum-rate, we will analyze the asymptotic distribution of mutual information 
and derive the closed-form expression for the outage probability to characterize reliability.


\section{Asymptotic Analysis of Reliability via Outage}
In \cite{ref.Jung2018lisul}, the mean and variance of the individual rate in an uplink LIS system is asymptotically derived. 
However, the distribution of the sum-rate must be analyzed in order to obtain the outage probability and evaluate the system reliability.
Deriving this distribution is not trivial and cannot be obtained directly from \cite{ref.Jung2018lisul}, because a sequence of individual rates, $\left[ {{R_1}, \cdots ,{R_K}} \right]$, is not identically distributed.
We begin with a characteristic of LIS systems in which
\begin{gather}
{S_k}/M^2 - {\bar p_k}/M^2 \xrightarrow[M \to \infty ]{} 0,\nonumber\\
{I_k}/{M^2}-{\bar \mu _{{I_k}}}/{M^2}\xrightarrow[M,K \to \infty ]{} 0,\nonumber
\end{gather}
where ${{\bar p}_k} = \frac{{{M^2}p_k^2}}{{16{\pi ^2}{L^4}}}$, ${p_k} = {\tan ^{ - 1}}( {{{{L^2}}}/({{{z_k}\sqrt {2{L^2} + z_k^2} }})} )$,
and ${\bar \mu _{{I_k}}}$ is a deterministic value depending on the correlation matrices and the locations of the devices \cite[Lemma 4]{ref.Jung2018lisul}.
Then, the asymptotic rate, ${\bar {R}}_k$, is obtained by the following lemma.

{\bf{{Lemma 1.}}} The asymptotic rate, ${\bar {R}}_k$, can be obtained by the function of a random variable $I_k$ as 
${\bar{R}}_k =  {{a_k} - {b_k}{I_k}}$,
where ${a_k} = \frac{{{\rho _k}{\bar p_k}( {1 - \tau_k^2} )}}{{{\rho _k}{\bar p_k}( {1 - \tau_k^2} ) {{+}} {{\bar \mu }_{{I_k}}}}}+\log (1+{\frac{{{\rho _k}{\bar p_k}( {1 - \tau_k^2} )}}{{{{\bar \mu }_{{I_k}}}}}} )$
and ${b_k} = \frac{{{\rho _k}{\bar p_k}( {1- \tau_k^2} )/{{\bar \mu }_{{I_k}}}}}{{{\rho _k}{\bar p_k}( {1- \tau_k^2} ) + \bar \mu _{{I_k}}}}$.
\begin{proof}
The detailed proof is presented in Appendix A.
\end{proof}

Given that ${\bar p_k}$ and ${{\bar \mu }_{{I_k}}}$ are deterministic values, $a_k$ and $b_k$ are also deterministic values.
Then, Lemma 1 shows that the distribution of ${\bar{R}}_k$ is exclusively determined by that of $I_k$,
and it allows us to readily obtain the distribution of ${\bar R}=\sum\nolimits_k {\bar R}_k$ by analyzing $I_k$ instead of ${\bar R}_k$ itself. 
From Lemma 1, we have ${\bar R} = \sum\nolimits_{k } {a_k-{b_k}{I_k}}$
and the second term $\sum\nolimits_{k } {{b_k}{I_k}} $ determines the distribution of $\bar R$.
By using the Lyapunov CLT \cite{ref.Ash2000probability}, we analyze the distribution of $\sum\nolimits_{k} {{b_k}{I_k}}$ and finally obtain the distribution of $R$, as follows.

{\bf{{Theorem 1.}}} For large $M$ and $K$, ${R}$ approximately follows $R \sim \mathcal{N}( {\bar\mu_{R}},{\bar\sigma_{R}^2}  ),$
where 
${\bar\mu_{R}} =\sum\nolimits_{k }\log ( 1 + \frac{{\rho _k}{\bar p_k}( {1 - \tau_k^2} )}{{\bar \mu }_{{I_k}}} )$ and 
${\bar\sigma_{R}^2}=\sum\nolimits_{k }\frac{{\bar \sigma _{{I_k}}^2\rho _k^2{\bar p_k}^2{{( {1 - \tau_k^2} )}^2}}}{{\bar \mu _{{I_k}}^2{{( {{{\bar \mu }_{{I_k}}} + {\rho _k}{\bar p_k}( {1 - {\tau_k ^2}} )} )}^2}}}$.

\begin{proof}
The detailed proof is presented in Appendix B.
\end{proof}


Theorem 1 shows that the sum-rate of an LIS system approximately follows a Gaussian distribution for large $M$ and $K$,
and its mean and variance can be obtained deterministically.
\begin{figure}[!ht]
\centering
\includegraphics[width=0.7\columnwidth] {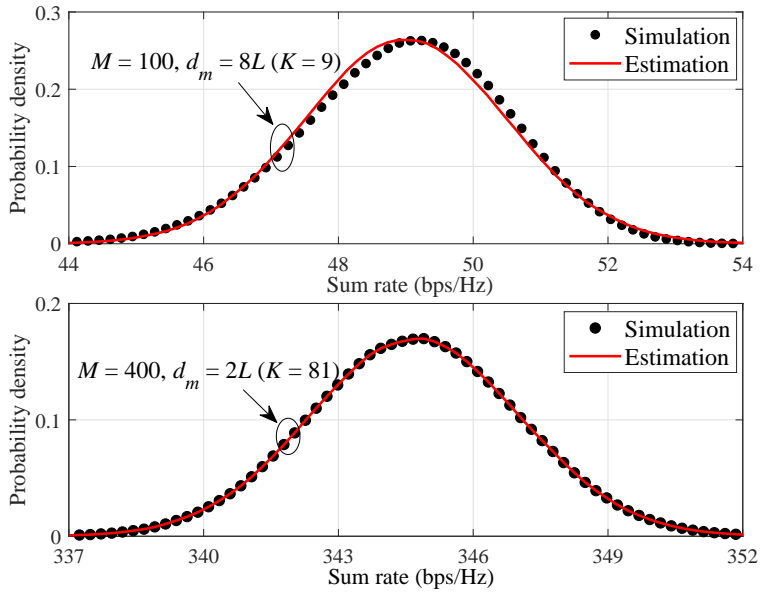}
\caption{PDFs of $R$ when $(M,d_m)$ is (top) $(100,8L)$ and (bottom) $(400,2L)$.}
\label{fig.2}
\end{figure}
This allows us to evaluate the performance of an LIS system in terms of outage probability, ergodic rate, reliability, and scheduling diversity, without extensive simulations.
Then, we can characterize the closed-form expression for the outage probability, defined as the probability when the instantaneous sum-rate falls below a certain threshold value.

{\bf{Corollary 1.}} For large $M$ and $K$, the outage probability of event $\left\{ {{R} < {R_{\rm{D}}}} \right\}$ is approximately obtained as follows:
\begin{equation}
P_{\rm{o}}=\Pr \left[ {{R} < {R_{\rm{D}}}} \right] = 1 - Q \big( {\frac{{{R_{\rm{D}}} - {\bar\mu_{R}}}}{{ {{\bar\sigma_{R}}} }}} \big),\nonumber
\end{equation}
where ${R_{\rm{D}}}$ is a desired rate when the probability of the outage event is $P_{\rm{o}}$, and $Q ( \cdot  )$ is a Gaussian Q-function.

Assume that $R_{\rm{D}}=\delta \cdot {\bar\mu_{R}}$ for $0< \delta \le 1$,
then we have ${{P}_{\rm{o}}}= Q \big( (1-\delta) \frac{\bar\mu_{R}}{\bar\sigma_{R}}\big)$.
As $M$ increases, ${\bar\mu_{R}}$ increases and converges to its bound,
and ${\bar\sigma_{R}}$ decreases, as proved in \cite{ref.Jung2018lisul}. 
Therefore ${\bar{P}_{\rm{o}}}$ decreases as $M$ increases.
On the basis of scaling law for $K$, ${\bar\mu_{R}}$ and ${\bar\sigma_{R}}$ follow $\mathcal{O}(K\log (1+\frac{1}{K}))=\mathcal{O}(1)$ and $\mathcal{O}(\frac{1}{\sqrt{K}})$, respectively.
Then, $\frac{\bar\mu_{R}}{\bar\sigma_{R}}$ increases with $\mathcal{O}(\sqrt{K})$ and finally ${\bar{P}_{\rm{o}}}$ also decreases, as $K$ increases.
Therefore, an LIS can provide \emph{reliable communication} in the presence of a large number of antennas and devices.
Moreover, since $\bar \mu _R$ and $\bar \sigma _R$ are deterministic values obtained from the correlation matrices and the locations of the devices,
we can estimate the outage probability of LIS without the need for extensive simulations.


\begin{figure}[!ht]
\centering
\includegraphics[width=0.7\columnwidth] {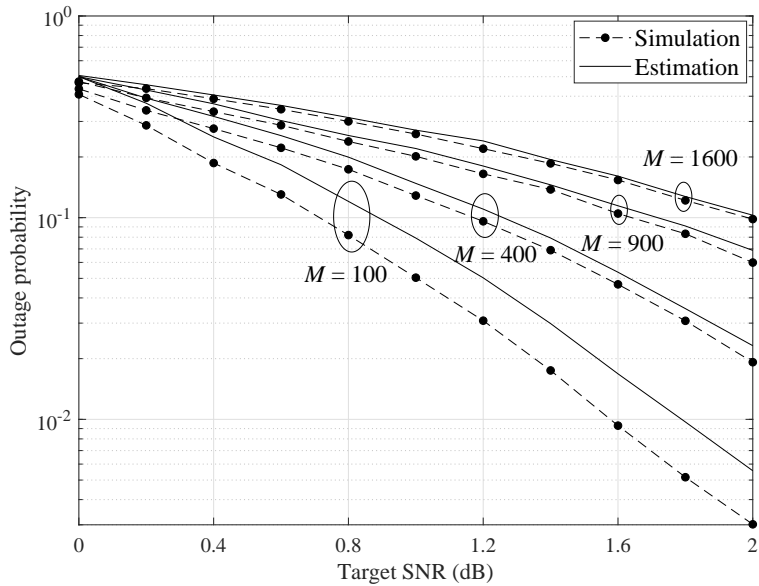}
\caption{Outage probabilities as a function of the target SNR when ${d_{\rm{m}}}=4L$ $(K=25)$.}
\label{fig.3}
\end{figure}

\section{Simulation Results and Analysis}
In this section, all simulations are statistically averaged over a large number of independent runs.
For our simulations, we assume that target SNR is $3$ dB, $\lambda=0.1$ m, $L = 0.25$ m, $\tau_k^2=0.5$, and $\beta_{\rm{PL}}=3.7$,
and the probability of LOS path and Rician factor are applied based on the 3GPP model \cite{ref.Jung2018lisul}.
We consider a scenario in which devices are located on a two-dimensional $xy$-plane at $z=1$ in parallel to the LIS within the range of $-2$ $\le x \le$ $2$ and $0$ $\le y \le$ $4$ (in meters)
and the distance between the adjacent devices is equally set to ${d_{\rm{m}}}$.
Therefore, a total of $81$, $25$, and $9$ devices are located in a two-dimensional rectangular lattice form when ${d_{\rm{m}}}=2L$, ${d_{\rm{m}}}=4L$, and ${d_{\rm{m}}}=8L$, respectively.
{For a massive MIMO system, we assume a single BS with $M$ antennas serving $K$ users based on a uniform linear array, as in \cite{ref.Jung2018lisul}.}

Fig. \ref{fig.2} shows the probability density function (PDF) of the instantaneous rate according to the independent channel realizations.
As proved in Theorem 1, the PDF of the sum-rate follows a Gaussian distribution and the accuracy of the Gaussian approximation improves as $M$ and $K$ increase.
As can be seen, the two PDFs are almost aligned with each other, and the approximation error is negligible when $M=400$ and ${d_{\rm{m}}}=2L$, corroborating the result of Theorem 1.


\begin{figure}[!ht]
\centering
\includegraphics[width=0.7\columnwidth] {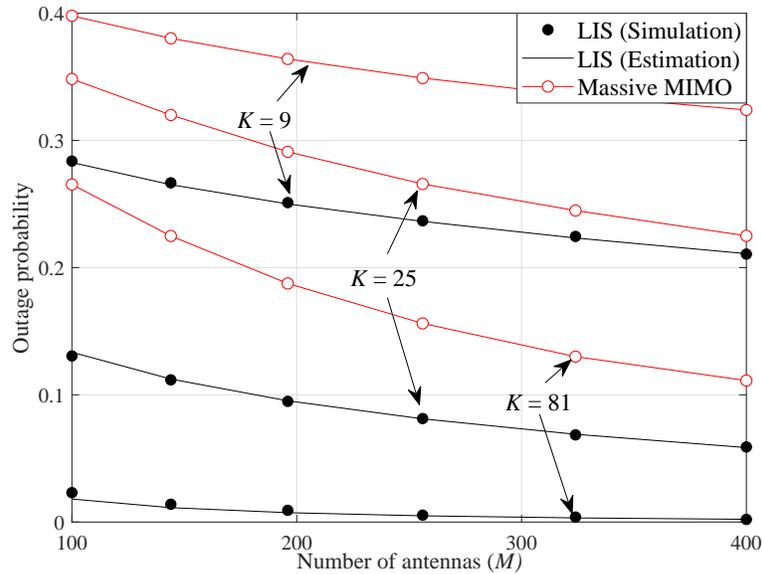}
\caption{{Outage probabilities as a function of $M$ when $\delta = 0.98$.}}
\label{fig.4}
\end{figure}

Fig. \ref{fig.3} and \ref{fig.4} compare the outage probabilities resulting from the simulations to the estimations from Corollary 1 as a function of the target SNR and $M$, respectively.
We assume that ${R_{\rm{D}}} = {\bar\mu_{R}}$ when the target SNR is $0$ dB in Fig. \ref{fig.3} and $\delta = 0.98$ in Fig. \ref{fig.4}.
{In both figures, the asymptotic results from Corollary 1 become close to the results of our simulations as $M$ increases,
verifying the accuracy of our analyses.}
As shown in Fig. \ref{fig.3}, even when $M=100$, the performance gap between the results of the asymptotic estimation and the simulation is less than $0.2$ dB in terms of the target SNR.
Moreover, since the noise component of an LIS system becomes negligible as $M$ increases \cite{ref.Jung2018lisul},
we can observe that the target SNR gradually lessens its effect on the outage probability as $M$ increases
{(e.g., when the target SNR is $2$ dB, the outage probabilities for $M=100$ and $1600$ are $3\cdot10^{-3}$ and $10^{-1}$, respectively)}.
{Furthermore, Fig. \ref{fig.4} shows that the outage probability decreases and eventually reaches zero as $M$ and $K$ increase.
In addition, the outage probabilities resulting from LIS are always lower than those resulting from massive MIMO system over the entire range of $M$, which shows that an LIS system can be more reliable than massive MIMO.}

\begin{figure}[!ht]
\centering
\includegraphics[width=0.7\columnwidth] {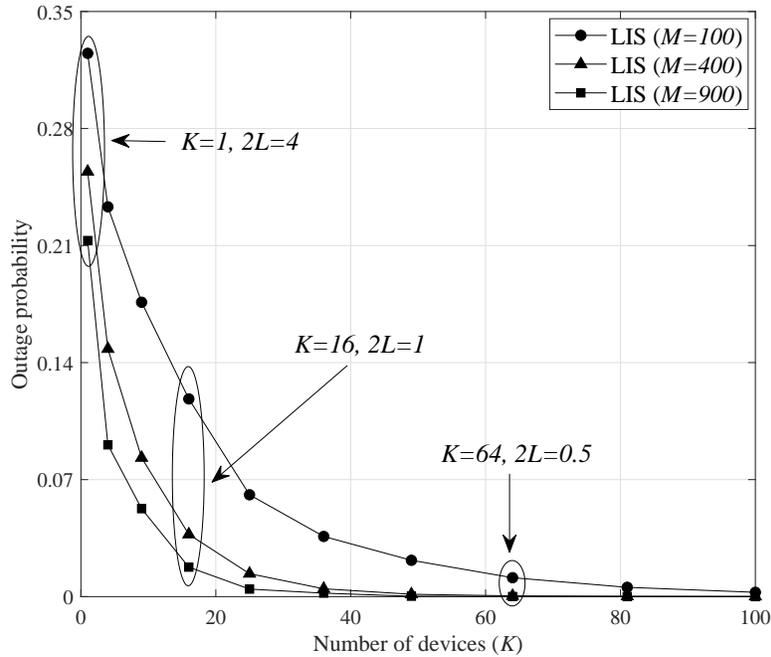}
\caption{{Outage probabilities as a function of $K$ when $\delta = 0.98$ and $A=16$.}}
\label{fig.5}
\end{figure}

{
Fig. \ref{fig.5} shows the impact of the surface-area of each LIS unit on the outage probability with a fixed total allocated area of an LIS (e.g., $A=16$ [m$^2$]).
Given a fixed $A$, $K$ increases as $L$ decreases and, hence, ${{P}_{\rm{o}}}$ decreases,
as shown in Fig. \ref{fig.5}.
}


\section{Conclusions}
In this paper, we have analyzed the sum-rate of an uplink LIS system asymptotically, under practical LIS environments. 
In particular, we have derived the distribution of the sum-rate by considering a practical LIS environment in which the interference can be generated by device-specific Rician fading with spatial correlation in presence of imperfect channel estimation.
Moreover, we have studied the outage probability of the sum-rate from the derived asymptotic distribution.
We have shown that the asymptotic results can accurately and analytically determine the performance of an LIS, including the distribution of the sum-rate and outage probability, without the need for a large number of simulations.
Simulation results have verified the results of the asymptotic analyses. 
These numerical results have shown that the analytic results were in close agreement with the results arising from extensive simulations,
and an LIS can provide reliable communication in terms of the outage probability. 

\section*{Appendix A\\Proof of Lemma 1}
We begin with the individual rate for device $k$ as ${{R}}_k = {\log }( {1 + \frac{{{\rho _k}{S_k}( {1 - \tau_k^2} )}}{{{I_k}}}}  )$.
Then, ${{R}}_k$ can be divided into two terms as ${{R}}_k = {{R_k^{\rm{L}}}} - {{R_k^{\rm{R}}}}$,
where ${{R_k^{\rm{L}}}} = {{{\log }}( {{\rho _k}{{S_k}}( {1 - \tau_k^2} )}+{{I_k}} )}$ and 
${{R_k^{\rm{R}}}} = {{{\log\rm{ } }} {{I_k}}}$.
Since ${{I_k}}/M^2-{\bar \mu _{{I_k}}}/M^2$ converges to zero as $M,K \to \infty$, 
we have the following series of equalities using the exponential function definition $e^x = \mathop {\lim }\limits_{n \to \infty } {( {1 + x/n} )^n}$:
\begin{align}
{{R_k^{\rm{L}}}} &= {\log }{\left( {1 + \frac{\left({{I_k} - {\bar \mu _{{I_k}}} }\right)/M^2}{\left({{{\rho _k}{{\bar p_k}}\left( {1 - \tau_k^2} \right)}  + {\bar \mu _{{{I_k}}}} }\right)/M^2}}\right)} +  c_k,\nonumber\\
{{{\bar R}_k^{\rm{L}}}}&=   {\frac{{{{I_k}} - {\bar \mu _{{{I_k}}}} }}{{{{\rho _k}{{\bar p_k}}\left( {1 - \tau_k^2} \right)}  + {\bar \mu _{{{I_k}}}} }}}  +  c_k,\nonumber
\end{align}
where $c_k = {\log}( {{{\rho _k}{{\bar p_k}}( {1 - \tau_k^2} )}  {+} {\bar \mu _{{I_k}}} } )$ and ${{R_k^{\rm{L}}}}-{{{\bar R}_k^{\rm{L}}}}\xrightarrow[\hspace{-2pt}M,K \to \infty \hspace{-2pt}]{} 0$.
Similarly, we have ${{\bar R}_k^{\rm{R}}} = {({{{I_k} -  {\bar \mu _{{I_k}}}}})/{ {\bar \mu _{{{I_k}}}} }} + {\log\rm{ }} {\bar \mu _{{I_k}}}$
and finally, 
we can obtain ${\bar{R}}_k = {{\bar R}_k^{\rm{L}}} - {{\bar R}_k^{\rm{R}}}$, which completes the proof.

\section*{Appendix B\\ Proof of Theorem 1}
Given the definition of ${I_k} = {{{\rho _k}\tau_k^2{X_k} + \sum\nolimits_{j \ne k}^K {{\rho _j}{Y_{jk}}}  + {Z_k}}}$, we make use of the following asymptotic convergence from \cite{ref.Jung2018lisul} pertaining to the distribution of ${I_{k}}$:
${I_k}-\sum\nolimits_{j \ne k}^K {{\rho _j}{Y_{jk}}}\xrightarrow[M \to \infty ]{} 0$. 
Then, $\sum\nolimits_{j \ne k}^K {{\rho _j}{Y_{jk}}}$ approximately determines the distribution of ${I_k}$ for large $M$.
Therefore, we have the following numerically accurate approximation for large $M$: $\sum\nolimits_{k} {{b_k}I_k} = \sum\nolimits_{k} {{b_k}\sum\nolimits_{j \ne k}^K {{\rho _j}{Y_{jk}}}}$.
Given a random variable $\sum\nolimits_{k } {{b_k}\sum\nolimits_{j \ne k}^K {{\rho _j}{Y_{jk}}}}$, we can observe that ${b_k}\sum\nolimits_{j \ne k}^K {{\rho _j}{Y_{jk}}}$ is a random variable independent across $k$ since ${{\boldsymbol{g}}_{jk}}$ and ${{\boldsymbol{e}}_k}$ are independent random vectors for different $k$.
Therefore, the following Lyapunov's condition should be satisfied for some $\delta>0$ and large $M$ to prove that $\sum\nolimits_{k } {{b_k}\sum\nolimits_{j \ne k}^K {{\rho _j}{Y_{jk}}}}$ is Gaussian distributed \cite{ref.Ash2000probability}:
\begin{equation}
\mathop {\lim }\limits_{K \to \infty } \frac{1}{{s_K^{2 {\rm{+}}\delta }}}\sum\limits_{k} {{\rm{E}}\left[ {{{\Big| {b_k}{\sum\nolimits_{j \ne k}^K {\rho _j}\left({{{Y_{jk}}} - {\bar \mu _{{Y_{jk}}}}}\right)} \Big|}^{2 {\rm{+}} \delta }}} \right]}  {\rm{=}} 0,\label{eq.A.CLTcon}
\end{equation}
where $s_K^2{=}\sum\nolimits_{k} {b_k^2}\bar \sigma _{{I_{k}}}^2$
and {
${\bar \mu _{{Y_{jk}}}}$ is asymptotic mean of ${Y_{jk}}$}.
Given that ${Y_{jk}}$ is a random variable correlated across $j$, 
we have
${Y_{jk}} = {s_{jk}}{| x_j |^2} + {| {\mu _{jk}^{\rm{L}}} |^2} + 2{\rm{Re}}( {\sqrt {{s_{jk}}} \mu _{jk}^{\rm{L}}{x^*_j}} ),$
where ${s_{jk}} = s_{jk}^{\rm{L}} + s_{jk}^{{\rm{N1}}} + s_{jk}^{{\rm{N2}}}$ and $x_j$ is a standard complex Gaussian random variable correlated across $j$.
{
The terms ${s_{jk}^{\rm{L}}},{s_{jk}^{\rm{N1}}},{s_{jk}^{\rm{N2}}},$ and ${{{| {{\mu _{jk}^{\rm{L}}}} |}^2}}$ are provided in \cite[Lemma 2]{ref.Jung2018lisul}, and ${s_{jk}}$ and ${{{| {{\mu _{jk}^{\rm{L}}}} |}^2}}$ are respectively increase with $\mathcal{O}(M)$ and $\mathcal{O}(M^2)$ as $M$ increases.}
We then calculate ${{{Y_{jk}}} - {\bar \mu _{{Y_{jk}}}}}$ in (\ref{eq.A.CLTcon}) as
${s_{jk}}( {{{| x_j |}^2} - 1} ) + 2{\mathop{\rm Re}\nolimits} ( {\sqrt {{s_{jk}}} \mu _{jk}^{\rm{L}}{x^*_j}} )$.
Given that $s_{jk}$ and $\sqrt {{s_{jk}}} \mu _{jk}^{\rm{L}}$ respectively increase with $\mathcal{O}(M)$ and $\mathcal{O}(M\sqrt{M})$,
we have ${{{Y_{jk}}} - {\bar \mu _{{Y_{jk}}}}}= 2{\mathop{\rm Re}\nolimits} ( {\sqrt {{s_{jk}}} \mu _{jk}^{\rm{L}}{x^*_j}} )$ for large $M$.
Then, we have the following inequality when $\delta=2$ in (\ref{eq.A.CLTcon}):
\begin{gather}
\mathop {\lim }\limits_{K \to \infty } \frac{{16\sum\nolimits_{k} {{{| {{b_k}} |}^4}{\rm{E}}\Big[ {{{\big| {\sum\nolimits_{j \ne k}^K {{\rho _j}} {\mathop{\rm Re}\nolimits} \big( {\sqrt {{s_{jk}}} \mu _{jk}^{\rm{L}}{x^*_j}} \big)} \big|}^4}} \Big]} }}{{{{\left( {\sum\nolimits_{k } {b_k^2} \bar \sigma _{{I_k}}^2} \right)}^2}}} \nonumber\\
\le \mathop {\lim }\limits_{K \to \infty } \frac{{16\sum\nolimits_{k} {{{| {{b_k}} |}^4}{\rm{E}}\Big[ {{{\Big| {\sum\nolimits_{j \ne k}^K {{\rho _j}} | x_j |\big| {\sqrt {{s_{jk}}} \mu _{jk}^{\rm{L}}} \big|} \Big|}^4}} \Big]} }}{{{{\left( {\sum\nolimits_{k } {b_k^2} \bar \sigma _{{I_k}}^2} \right)}^2}}}.\label{eq.A.Ineq1}
\end{gather}
The expectation term in the numerator of (\ref{eq.A.Ineq1}) is bounded as
\begin{align}
&{\rm{E}}\left[ {{{\Big| {\sum\nolimits_{j \ne k}^K {{\rho _j}| {{x_j}} |} | {\sqrt {{s_{jk}}} \mu _{jk}^{\rm{L}}} |} \Big|}^4}} \right]
\mathop  \le \limits_{\left( \rm{a} \right)} {\rm{E}}\left[ {{{\big( { d_k\sum\nolimits_{j \ne k}^K {\rho _j^2{{| {{x_j}} |}^2}} } \big)}^2}} \right]\nonumber\\
 &= {d_k^2}\Big( {{\rm{Var}}\Big[ {\sum\nolimits_{j \ne k}^K {\rho _j^2{{| {{x_j}} |}^2}} } \Big]{\rm{+}}{\rm{ E}}{{\Big[ {\sum\nolimits_{j \ne k}^K {\rho _j^2{{| {{x_j}} |}^2}} } \Big]}^2}} \Big)\nonumber\\
 &\mathop  \le \limits_{( \rm{b} )} {d_k^2}\Big( {\sum\nolimits_{j \ne k}^K {\rho _j^4} {\rm{+}}\sum\nolimits_{i,j \ne k:i \ne j}^K {\rho _i^2\rho _j^2} {\rm{+}}{{\Big( {\sum\nolimits_{j \ne k}^K {\rho _j^2} } \Big)}^2}}
 \Big), 
\label{eq.A.Ineq2}
\end{align}
where $d_k = \sum\nolimits_{j } {{s_{jk}}{{| {\mu _{jk}^{\rm{L}}} |}^2}}$, (a) results from Cauchy--Schwarz inequality, and (b) results from the covariance inequality \cite{ref.Mukhopadhyay2000probability}:
\begin{align}
&\sum\nolimits_{i,j \ne k:i \ne j}^K {\rho _i^2\rho _j^2} {\rm{Cov}}\big[ {{{| {{x_i}} |}^2},{{| {{x_j}} |}^2}} \big]\nonumber\\
&\mathop  \le \limits_{\left( \rm{c} \right)}\sum\nolimits_{i,j \ne k:i \ne j}^K {\rho _i^2\rho _j^2} \sqrt {{\rm{Var}}\big[ {{{| {{x_i}} |}^2}} \big]{\rm{Var}}\big[{{{| {{x_j}} |}^2}} \big]}= \sum\limits_{i,j \ne k:i \ne j}^K {\rho _i^2\rho _j^2}.\nonumber
\end{align}

\noindent The covariance inequality (c) always holds because the covariance matrix of a random vector $\boldsymbol{x}$ is a real symmetric matrix \cite{ref.Jung2018lisul}, and it is therefore positive-definite
when we define ${\boldsymbol{x}} = \left[ {{x_1}, \cdots ,{x_{k - 1}},{x_{k + 1}}, \cdots ,{x_K}} \right]^{\rm{T}}$.
We then determine the scaling laws of (\ref{eq.A.Ineq2}) according to $M$ and $K$.
Since ${{s_{jk}}{{\big| {\mu _{jk}^{\rm{L}}} \big|}^2}}$ increases with $\mathcal{O}(M^3)$, 
$d_k^2$ increases with $\mathcal{O}(K^2M^6)$ and then (\ref{eq.A.Ineq2}) increases with $\mathcal{O}(K^4 M^6)$ as $M,K\to\infty$.
{
Further, given that $\bar \mu _{{I_{k}}}$ and $\bar \sigma _{{I_{k}}}^2$ respectively increase with $\mathcal{O}(KM^2)$ and $\mathcal{O}(K^2M^3)$ as proved in \cite{ref.Jung2018lisul},
$b_k$ in (\ref{eq.A.Ineq1}) decreases with ${\mathcal{O}(\frac{1}{K^2M^4})}$ as $M,K\to\infty$.}
Then, the numerator and denominator of the upper bound of (\ref{eq.A.Ineq1}) decrease with ${\mathcal{O}(\frac{1}{K^3M^{10}})}$ and ${\mathcal{O}(\frac{1}{K^2M^{10}})}$, respectively, as $M,K\to\infty$.
Therefore, the upper bound of (\ref{eq.A.Ineq1}) decreases with ${\mathcal{O}(\frac{1}{K})}$ and eventually reaches zero as $K\to\infty$.
In conclusion, (\ref{eq.A.CLTcon}) is satisfied and $\sum\nolimits_{k} {{b_k}{I_k}}$ approximately follows a Gaussian distribution.
Given that ${b_k}{I_k}$ is a random variable independent across $k$, we can ultimately obtain the distribution of $\sum\nolimits_{k } {{b_k}{I_k}}$ for large $M$ and $K$, as follows:
$\sum\nolimits_{k} {{b_k}{I_k}}\sim \mathcal{N}\left( {\sum\nolimits_{k} {{b_k}{{\bar \mu }_{{I_k}}}} ,\sum\nolimits_{k } {b_k^2\bar \sigma _{{I_k}}^2} } \right).$
For large $M$ and $K$, we have the following numerically accurate approximation: $R = \sum\nolimits_{k }a_k{-} {{b_k}{I_k}}$. 
Therefore,
$R\sim \mathcal{N}\left( {\sum\nolimits_{k} {a_k{-}{b_k}{{\bar \mu }_{{I_k}}}} ,\sum\nolimits_{k } {b_k^2\bar \sigma _{{I_k}}^2} } \right),$
which completes the proof.

\bibliographystyle{IEEEtran}
\bibliography{IEEEabrv,myBiB_response}
\end{document}